\documentstyle[12pt,a4wide]{article}

\newcommand{\be}[1]{\begin{equation}\label{#1}}
\newcommand{\ee}{\end{equation}}
\newcommand{\ba}[1]{\begin{eqnarray}\label{#1}}
\newcommand{\ea}{\end{eqnarray}}
\newcommand{\R}{ \mbox{\rm I$\!$R} }

\begin{document}

%*****************************************************************
%*****************************************************************

\centerline{\Large\bf MULTIDIMENSIONAL CLASSICAL AND}
\bigskip
\centerline{\Large\bf QUANTUM WORMHOLES IN  MODELS}
\bigskip
\centerline{\Large\bf  WITH COSMOLOGICAL CONSTANT}
\bigskip

%%%%%%%%%%%%%%%%%%%%%%%%%%%%%%%%%%%%%%%%%%%%%%%%%%%%%%%%%%%%%%%%%%%%%%%

\bigskip

\bigskip
\centerline{ \bf U. BLEYER
\footnote{E-mail address: UBleyer@aip.de. This work was supported
by KAI e.V. grant WIP-016659},
V. D. IVASHCHUK$^{2}$, V. N. MELNIKOV\footnote{Permanent address:
Center for Surface and Vacuum Research, 8
Kravchenko str., Moscow, 117331, Russia,
e-mail: mel@cvsi.uucp.free.msk.su.
This work was supported in part by DFG grant 436
RUS - 113-7-2
}
}
{\small
\centerline{WIP-Gravitationsprojekt, Universit\"at Potsdam }
\centerline{An der Sternwarte 16 }
\centerline{ D-14482 Potsdam, Germany}
}
\bigskip
\centerline{and}
\bigskip
\centerline{\bf A. ZHUK
\footnote{Permanent address: Department of Physics,
University of Odessa,
2 Petra Velikogo,
Odessa 270100, Ukraine.
This work was supported in part by DAAD and by DFG grant 436
RUS - 113-7-1}
}
{\small
\centerline{WIP-Gravitationsprojekt, Universit\"at Potsdam }
\centerline{An der Sternwarte 16 }
\centerline{ D-14482 Potsdam, Germany}
\centerline{and}
\centerline{Fachbereich Physik, Freie Universit\"at Berlin}
\centerline{Arnimallee 14}
\centerline{ D-14195 , Germany}
}
%*****************************************************************
%*****************************************************************
\bigskip

\begin{abstract}
\noindent
A multidimensional cosmological model  with space-time consisting of
$n (n \ge 2)$ Einstein spaces $M_{i}$ is investigated in the presence
of a cosmological constant $\Lambda$ and a homogeneous minimally
coupled scalar field $\varphi(t)$ as a matter source. Classical and
quantum wormhole solutions are obtained for $\Lambda < 0$ and all
$M_{i}$ being Ricci-flat. Classical wormhole solutions are also found
for $\Lambda < 0$ and only one of the $M_{i}$ being Ricci-flat for the
case of spontaneous compactification of the internal dimensions with
fine tuning of parameters.
\end{abstract}

%%%%%%%%%%%%%%%%%%%%%%%%%%%%%%%%%%%%%%%%%%%%%%%%%%%%%%%%%%%%%%%%%%
%%%%%%%%%%%%%%%%%%%%%%%%%%%%%%%%%%%%%%%%%%%%%%%%%%%%%%%%%%%%%%%%%%
\section{INTRODUCTION}
\setcounter{equation}{0}
%%%%%%%%%%%%%%%%%%%%%%%%%%%%%%%%%%%%%%%%%%%%%%%%%%%%%%%%%%%%%%%%%

In quantum cosmology instantons, solutions of the classical
Einstein equations in Euclidean space, play an important role
giving the main contributions to the path integral \cite{1}. Among
them classical wormholes are of special interest, because they are
connected with processes changing the topology of the models
\cite{2,3}. We remind that classical wormholes are usually Riemannian
metrics consisting of two large regions joined by a narrow throat
(handle). They exist for special types of matter \cite{2} - \cite{4}
and do not exist for pure gravity. In quantum cosmology it is
generally assumed that on Planck scale processes with topology changes
should take place. For this reason Hawking and Page
\cite {5} introduced the notion of quantum wormholes as a quantum
extension of the classical wormhole paradigma. They proposed to regard
quantum wormholes  as solutions of the Wheeler-DeWitt (WDW) equation
with the following boundary conditions:

(i) the wave function is
exponentially damped for large spatial geometry,

(ii) the wave
function is regular when the spatial geometry degenerates.

The first condition expresses the fact that space-time should be
Euclidean at spatial infinity. The second condition should reflect the
fact that space-time is nonsingular when spatial geometry degenerates.
For example, the wave function should not oscillate an infinite number
of times.

The given approach extends the number of objects which can be treated
as wormholes \cite{5,6}.

We believe that for the description of quantum gravitational processes
at high energies the multidimensional approach is more adequate. Modern
theories of unified physical interactions use ideas of hidden (or
extra) dimensions. In order to study different phenomena at early
stages of the universe one should use these theories or at any rate
models keeping their main characteristics. But more reliable
conclusions may be done only on the basis of exact solutions which are
usually obtained in rather simple cases.

Therefore, at the beginning we consider a cosmological model with $n$
$(n > 1)$ Einstein spaces containing a massless minimally
coupled scalar field and a cosmological constant $\Lambda$. The gauge
covariant form of the WDW equation was proposed in
\cite{7}. This model is integrable in the case with only one of the
Einstein spaces being not Ricci-flat and vanishing
cosmological constant. The general properties of this particular model
were investigated in \cite{8} while classical as well as quantum
wormhole solutions were found for different models in \cite{9} -
\cite{11a}. (In \cite{12} particular integrable cases with milticomponent
perfect fluid were considered).

The present paper is devoted to the case of nonzero cosmological
constant. For the model with one space of positive constant curvature
in four dimensional space-time with cosmological constant and axionic
matter (which is equivalent to a free minimally coupled scalar field,
see, for example, \cite{2,3} and the paper by Brown et al. in \cite{4})
classical wormhole solutions were obtained in \cite{3,11b}. In the case
of four dimensional space-time with nontrivial topology $\R \times
S^{1} \times S^{2}$ and non-zero cosmological constant this type of
solution exists, too \cite{12a}.

Here, we investigate the case with at least one of the spaces, say
$M_{1}$ being Ricci-flat. If all the other spaces $M_{i}, i = 2,
\dots, n$ are also Ricci-flat this model is fully integrable in the
classical \cite{11a} as well as in the quantum cases \cite{7}. For
$\Lambda < 0$ a family of
quantum wormhole solutions with  continuous and discrete spectra
exist. Classical wormholes can be found in this case only in the
presense of a scalar field. In the presence of a  scalar field
classical wormhole solutions exist also for another particular case with
fine tuning of the parameters of the model, if $\Lambda < 0$ and all
$M_{i}, i = 2, \dots, n$ are not Ricci-flat and have the same sign of
the curvature. In this case only $M_{1}$ has a dynamical behaviour and
is considered as our external space. All the other internal spaces
$M_{i},  i = 2, \dots, n$ are freezed with fixed scale factors
$a_{(0)i}$ which are fine tuned to values determined by the
cosmological constant. This type of solutions belongs to the class of
models with spontaneous compactification. In the case of models
without a cosmological constant and with only one non-Ricci-flat
factor space solutions with spontaneous compactification were also
found in \cite{13}.

We would like to note that solutions of the WDW equation in four
dimensional models with $\Lambda \ne 0$ and with a conformal scalar field
were first obtained in \cite{14} and \cite{15} respectively (see also
\cite{16}). They include possibly the first quantum wormhole type
solutions in four dimensions as well as DeWitt's solution for the
Friedman universe with dust (1967). Vacuum quantum cosmological
solutions in four dimensions may be found in \cite{17b}.
The path integral approach to quantum cosmology \cite{1}
for models with cosmological constant in four and five dimensions
with nontrivial topology was developed in \cite{17}.

The paper is organized as follows. In section 2  the general
description of the models considered is given. In section 3 classical
and   quantum
wormholes are obtained for all spaces being Ricci-flat. In section 4
classical wormholes are considered in the model with spontaneous
compactification of extra dimensions. Conclusions and an extensive list
of references complete the paper.

%%%%%%%%%%%%%%%%%%%%%%%%%%%%%%%%%%%%%%%%%%%%%%%%%%%%%%%%%%%%%%%%%%%%
%%%%%%%%%%%%%%%%%%%%%%%%%%%%%%%%%%%%%%%%%%%%%%%%%%%%%%%%%%%%%%%%%%%%
\section{GENERAL DESCRIPTION \newline  OF THE MODEL}
\setcounter{equation}{0}
%%%%%%%%%%%%%%%%%%%%%%%%%%%%%%%%%%%%%%%%%%%%%%%%%%%%%%%%%%%%%%%%%%%%

The metric of the model
\begin{equation}\label{1}
g=-exp[2{\gamma}(\tau)]d\tau \otimes d\tau +
\sum_{i=1}^{n} exp[2{\beta^{i}}(\tau)]g^{(i)},
\end{equation}
is defined on the  manifold
\begin{equation}\label{2}
M = \R \times M_{1} \times \ldots \times M_{n},
\end{equation}
where the manifold $M_{i}$ with the metric $g^{(i)}$ is an
Einstein space of dimension $d_{i}$, i.e.
\begin{equation}\label{3}
{R_{m_{i}n_{i}}}[g^{(i)}] = \lambda^{i} g^{(i)}_{m_{i}n_{i}},
\end{equation}
$i = 1, \ldots ,n $; $n \geq 2$.
The total dimension of the space-time $M$ is $D = 1 +
\sum_{i=1}^{n} d_{i}$.

Here we investigate the general model with cosmological constant
$\Lambda$ and a homogeneous minimally coupled field
${\varphi}(t)$ with a potential ${U}(\varphi)$.

The action of the model is adopted in the following form
%8
\begin{equation}\label{8}
S =  \frac{1}{2}
\int d^{D}x \sqrt{|g|}
\{ {R}[g] -   \partial_{M} \varphi \partial_{N} \varphi g^{MN}
- 2{U}(\varphi) - 2 \Lambda \} + S_{GH},
\end{equation}
where $R[g]$ is the scalar
curvature of the metric $g = g_{MN} dx^{M} \otimes dx^{N}$ and
$S_{GH}$ is the standard Gibbons-Hawking boundary term \cite{17a}. The
field equations, corresponding to the action (\ref{8}), for the
cosmological metric (\ref{1}) in the harmonic time gauge
$\gamma  \equiv \sum_{i=1}^{n} d_{i} \beta^{i} $ are equivalent
to the Lagrange equations, corresponding to the Lagrangian
\begin{equation}\label{9}
L = \frac{1}{2}\sum_{i,j= 1}^{n}
( G_{ij}\dot{\beta}^{i}\dot{\beta}^{j} +
\dot{\varphi}^{2})  - V ,
\end{equation}
with the energy constraint imposed
%10
\begin{equation}\label{10}
E = \frac{1}{2} \sum_{i,j= 1}^{n}
( G_{ij}\dot{\beta}^{i}\dot{\beta}^{j}
+ \dot{\varphi}^{2})  + V = 0.
\end{equation}
Here, the overdot denotes differentiation with respect to the
harmonic time $\tau$. The components of the minisuperspace metric
read
%11
\begin{equation}\label{11}
G_{ij}= d_{i}\delta_{ij}- d_{i}d_{j} \end{equation}
and the potential is given by
%12
\be{12}
V = V(\beta, \varphi) = \exp(2\sum_{i=1}^{n} d_i \beta ^i)
\left[ -\frac{1}{2} \sum_{j=1}^{n} \theta_{j} e^{-2\beta ^j}+
{U}(\varphi )+ \Lambda \right] \ ,
\end{equation}
where $\theta_i = \lambda^id_i$. If the $M_i$ are spaces of constant
curvature, then $\theta_i$ may be normalized in such a way that
$\theta_i= k_id_i(d_i-1), ~k_i=\pm1,0$.
We may also consider the generalization of the model with the
potential (\ref{12}) modified by the substitution
%13
\be{13}
{U}(\varphi )  \mapsto  {\tilde U}(\varphi, \beta).
\ee
This gives us the possibility to investigate models with an arbitrary
scalar field potential ${\tilde U}(\varphi, \beta) \equiv U(\varphi)$
as well as (for $\varphi = $const) models  with an arbitrary potential
${\tilde U}(\varphi, \beta) \equiv U(\beta)$.  Effective potentials of
the form $U(\beta)$ may have their origin in an ideal fluid matter
source. In special cases the general form  ${\tilde U}(\varphi, \beta)
$ of the potential leads us to new integrable models. An example of
this kind of potential will be presented.

With the general potential ${\tilde U}(\varphi, \beta)$ the equations
of motion are
%14
$$
-d_i\ddot \beta ^i+ d_i \sum_{k=1}^{n} d_{k} \ddot \beta
^k+e^{2\sum_{k=1}^nd_k\beta ^k}\left[ \left( d_i-1\right) \theta
_ie^{-2\beta ^i}+d_i\sum_{k\neq i}\theta _{k\,}e^{-2\beta ^k} \right.
$$
\be{14}
\left.
-\frac{\partial \tilde U}{\partial \beta ^i}-
2d_i\tilde U-2d_i\Lambda \right] =0\ ,\qquad i=1,\ldots ,n,
\end{equation}
%15
\be{15}
\ddot \varphi \,+\frac{\partial \tilde U}{\partial \varphi }%
\exp \left(2\sum_{i=1}^{n} d_i\beta ^i \right) = 0 .
\end{equation}
with the constraint (\ref{10}).
%16
%\be{16}
%\frac 12 \left( \sum_{i,j=1}^{n} G_{ij} \dot \beta ^i\dot \beta ^j+
%\dot \varphi
%^2\right) +V = 0 .
%\end{equation}

At the quantum level the constraint (\ref{10}) is modified into the WDW
equation (see \cite{7})
%17
\be{17}
\left[
\frac 12 \left( G^{ij} \frac{\partial}{\partial\beta^{i}}
\frac{\partial}{\partial\beta^{j}} +
\frac{\partial^{2}}{\partial\phi^{2}}  \right)  - V(\beta, \varphi)
\right] \Psi(\beta, \varphi) = 0 ,
\ee
where $\Psi={\Psi}(\beta, \varphi)$ is the wave function of the universe,
$V$ is the potential (\ref{12}) and
%18
\be{18}
G^{ij} = \frac{\delta^{ij}}{d_{i}}+\frac{1}{2-D}
\end{equation}
are the components of the matrix inverse to the matrix $(G_{ij})$
(\ref{11}). The minisuperspace metric $G = G_{ij}dx^{i}\otimes dx^{i}$
(\ref{11}) was diagonalized in \cite{7,17c}
%19
\be{19}
G= -dz^{0} \otimes dz^{0} + \sum_{i=1}^{n-1}dz^{i}\otimes dz^{i},
\end{equation}
where
%20
\ba{20}
&& z^{0}= q^{-1} \sum_{j=1}^{n}d_{j}\beta^{j}, \nonumber \\
&& z^{i} = [d_{i}/ \Sigma_{i} \Sigma_{i+1}]^{1/2} \sum_{j=i+1}^{n}
d_{j}(\beta^{j}-\beta^{i}),
\end{eqnarray}
$i=1, \ldots ,n-1$, where
%21
\be{21}
q=[(D-1)/(D-2)]^{1/2}, \qquad \Sigma_{i} = \sum_{j=i}^{n}d_{j}.
\end{equation}
The WDW equation (\ref{17}) takes in variables (\ref{20}) the
following form
%22
\be{22}
\left[-\frac{\partial}{\partial z^{0}} \frac{\partial}{\partial z^{0}}
+ \sum_{i=1}^{n-1} \frac{\partial}{\partial z^{i}}
\frac{\partial}{\partial z^{i}} + \frac{\partial^{2}}{\partial\varphi ^{2}}
- 2 V(z, \varphi)\right]\Psi=0 .
\end{equation}

\bigskip

%%%%%%%%%%%%%%%%%%%%%%%%%%%%%%%%%%%%%%%%%%%%%%%%%%%%%%%%%%%%%%%%%%%%
%%%%%%%%%%%%%%%%%%%%%%%%%%%%%%%%%%%%%%%%%%%%%%%%%%%%%%%%%%%%%%%%%%%%
\section{WORMHOLES FOR  \newline RICCI-FLAT SPACES  }
\setcounter{equation}{0}
%%%%%%%%%%%%%%%%%%%%%%%%%%%%%%%%%%%%%%%%%%%%%%%%%%%%%%%%%%%%%%%%%%%%

In this chapter we consider the Ricci-flat case  $(\theta_i=
\lambda_{i}d^i= 0$, $i = 1, \ldots, n)$, with $\Lambda \neq 0$ and
${U}(\varphi) = 0$. If the $M_i$ are internal spaces they should be
compact. This compactness is a necessary condition also for the
Hartle-Hawking boundary condition (see below). The compactness of
Ricci-flat spaces may be achieved by appropriate boundary conditions.
The d-dimensional tore is the simplest example.

\subsection{Classical solutions}

In the considered case the Lagrangian (\ref{9}) may be written in the
following form
\begin{equation}\label{23}
L = \frac{1}{2} \sum_{I,J= 1}^{n+1}
 G_{IJ}\dot{\beta}^{I}\dot{\beta}^{J}
 - \Lambda \exp( \sum_{I= 1}^{n+1} u_{I} \beta^{I}),
\end{equation}
where $\beta^{n+1} = \varphi$, $u_{i} = 2d_{i}, i=1, \ldots, n$, $u_{n+1} =0$
and
\begin{equation}\label{24}
\left( G_{IJ} \right) =         \left(
                              \begin{array}{cc}
                                {G_{ij}} &  0 \\
                                   0   &  1
                               \end{array}
                          \right)
\end{equation}
(the matrix $(G_{ij})$ is defined in eq. (\ref{11})). We consider the
coordinates $(z^{A}) = (z^{a}, z^{n} = \beta^{n+1} = \varphi)$,
where  $z^{a} $,  $a  = 0, \ldots, n-1$, are defined in (\ref{20}).
It is clear that
\begin{equation}\label{25}
z^{A} =  \sum_{I= 1}^{n+1} V_{I}^{A} \beta^{I},
\end{equation}
$A  = 0, \ldots, n$, where
\begin{equation}\label{26}
\left(V^{A}_{I}\right) = \left(
                              \begin{array}{cc}
                                {V^{a}_{i}} &  0 \\
                                   0      &  1
                              \end{array}
                          \right)
\end{equation}
 and the matrix ($V^{a}_{i}$) is defined in (\ref{20}).
 This introduced matrix diagonalizes the minisuperspace metric
 \begin{equation}\label{27}
G_{IJ} = \sum_{A,B = 0}^{n} \eta_{AB} V_{I}^{A} V^{B}_{J},
\end{equation}
$I,J  = 1, \ldots, n+1$ and $(\eta_{AB}) = (\eta^{AB}) = \mbox{diag}(-1,
+1, \ldots, +1)$.

In the coordinates (\ref{25}) the Lagrangian (\ref{23}) reads ($q$ is
defined in (\ref{21}))
\begin{equation}\label{28}
L = \frac{1}{2} \sum_{A,B= 0}^{n}
 \eta_{AB} \dot{z}^{A}\dot{z}^{B}
 - \Lambda \exp( 2q z^{0}).
\end{equation}
The Lagrange equations for the Lagrangian (\ref{28})
\begin{eqnarray}\label{29}
&&-\ddot{z}^{0} + 2q \Lambda \exp(2q z^{0}) = 0, \\
&&\ddot{z}^{A} = 0, \qquad A  = 1, \ldots, n, \label{30}
\end{eqnarray}
with the energy constraint
\begin{equation}\label{31}
E = \frac{1}{2} \sum_{A,B= 0}^{n}
 \eta_{AB} \dot{z}^{A}\dot{z}^{B}
 + \Lambda \exp( 2q z^{0}) = 0
\end{equation}
can be readily solved. First integrals of (\ref{30}) are
\be{31a}
\dot z^A = p^A, \qquad A = 1, \ldots, n,
\ee
where $p^A$ are arbitrary constants of integration. Then the
constraint (\ref{31}) may be rewritten
\be{31b}
- \frac12 (\dot z^0)^2 + {{\cal E}} + \Lambda e^{2qz^0} = 0
\ee
with
\begin{equation}\label{31c}
2{{{\cal E}}} =  \sum_{A=1}^{n} (p^{A})^{2}.
\end{equation}
We obtain the following solution
\begin{equation}\label{32}
z^{A} = p^{A}\tau + q^{A}, \qquad         A=1, \ldots , n,
\end{equation}
where $p^{A}$ and $q^{A}$ are constants and
\begin{eqnarray}\label{33}
2qz^{0} & =
& {\ln[{{{\cal E}}} / \{\Lambda \sinh^{2}(q \sqrt{2{{\cal
E}}}(\tau-\tau_{0}))\}],}
\qquad {{{\cal E}} \neq 0, ~\Lambda > 0,}   \\
& = & {\ln[1/\{2q^{2} \Lambda (\tau - \tau_{0})^{2}\}],} \qquad
{ {{{\cal E}}}=0, ~\Lambda > 0,} \label{34}    \\
& = & { \ln[-{{{\cal E}}}/ \{\Lambda \cosh^{2}(q \sqrt{2{{{\cal E}}}}(\tau
- \tau_{0}))\}],}
\qquad {{{{\cal E}}} > 0, ~\Lambda < 0,} \label{35}
\end{eqnarray}
Here $\tau_{0}$ is an arbitrary constant.

%%%%%%%%%%%%%%%%%%%%%%%%%%%%%%%%%%%%%%%%%%%%%%%%%%%%%%%%%%%%
\subsection{Kasner-like parametrization}
%%%%%%%%%%%%%%%%%%%%%%%%%%%%%%%%%%%%%%%%%%%%%%%%%%%%%%%%%%%%%

First we consider the case
${{{\cal E}}} > 0$. In this case the relations (\ref{33}) and
(\ref{35})
may be written in the following form
\begin{equation}\label{37}
2qz^{0} =
\ln[{{{\cal E}}} /\{ |\Lambda| {f^{2}_{\delta}}(q \sqrt{2{{{\cal E}}}}(\tau-
\tau_{0}))\}],
\end{equation}
where $\delta \equiv  \Lambda / |\Lambda| = \pm 1$ and
\begin{eqnarray}\label{38}
{f_{\delta}}(x) \equiv \frac{1}{2} (e^{x} - \delta e^{-x})
& = &\sinh x, \qquad \delta = +1, \nonumber \\
& = &\cosh x, \qquad \delta = -1.
\end{eqnarray}
We introduce a new time variable by the relation
\begin{equation}\label{39}
t = \frac{T}{\sqrt{ \delta}} \ln \frac
{\exp(q \sqrt{2{{{\cal E}}}}(\tau - \tau_{0})) + \sqrt{ \delta}}
{\exp(q \sqrt{2{{{\cal E}}}}(\tau -\tau_{0})) -
\sqrt{ \delta}},
\end{equation}
where
\begin{equation}\label{40}
T \equiv [(D - 2)/ 2|\Lambda|(D-1)]^{1/2}.
\end{equation}
It is not difficult to verify that the following relations take
place
\begin{eqnarray}\label{41}
\sinh(t \sqrt{\delta}/T)/ \sqrt{ \delta} & = &
1/{f_{\delta}}(q \sqrt{2{{{\cal E}}}}(\tau - \tau_{0})), \\
\tanh(t \sqrt{\delta}/2T)/ \sqrt{ \delta} & = &
\exp(-q \sqrt{2{{{\cal E}}}}(\tau - \tau_{0})), \label{42} \\
dt & = & - T q \sqrt{2{{{\cal E}}}} d\tau/
{f_{\delta}}(q \sqrt{2{{{\cal E}}}}(\tau - \tau_{0})). \label{43}
\end{eqnarray}
Now, we introduce the following dimensionless parameters
\begin{equation}\label{44}
\bar{\alpha}^{I} \equiv - \sum_{A=1}^{n}
\bar{V}^{I}_{A} p^{A}/ q \sqrt{ 2 {{{\cal E}}}},
\end{equation}
where $(\bar{V}^{I}_{A})  = (V^{A}_{I})^{-1}$
(see (\ref{25})). It is clear that
\begin{equation}\label{45}
\left(\bar{V}_{A}^{I} \right) = \left(
                              \begin{array}{cc}
                                \bar{V}_{a}^{i} &  0 \\
                                   0            &  1
                               \end{array}
                          \right)
\end{equation}
where $(\bar{V}_{a}^{i}) = (V^{a}_{i})^{-1}$. The relation
(\ref{44}) is equivalent to the following relations
\begin{eqnarray}\label{46}
\bar{\alpha}^{i} & = & - \sum_{a=1}^{n-1}
\bar{V}^{i}_{a} p^{a}/ q \sqrt{ 2 {{{\cal E}}}}, \quad i = 1, \ldots, n, \\
&&\bar{\alpha}^{n+1} = -  p^{n}/ q \sqrt{ 2 {{{\cal E}}}}. \label{47}
\end{eqnarray}
It follows from eq. (\ref{27}), that
\begin{equation}\label{48}
\bar{V}_{A}^{I} = \sum_{J=1}^{n+1} \sum_{B=0}^{n}
G^{IJ} V_{J}^{B} \eta_{BA}
\end{equation}
where
\begin{equation}\label{49}
\left(G^{IJ} \right) \equiv  \left(G_{IJ} \right)^{-1} =
                             \left(
                              \begin{array}{cc}
                                {G^{ij}} &  0 \\
                                   0   &  1
                               \end{array}
                          \right).
\end{equation}

{}From (\ref{18}) and (\ref{48}) we have
\begin{equation}\label{51}
\bar{V}^{i}_{0} = - G^{ij} u_{j}/ 2q = (q(D-2))^{-1};
\qquad \bar{V}^{n+1}_{0} = 0.
\end{equation}
Using the relations (\ref{32}), (\ref{37}), (\ref{41})-(\ref{43}),
(\ref{44}) and  (\ref{51})
we get the folowing expressions for the solution of field
equations
\begin{eqnarray}\label{52}
g &=&- dt \otimes dt +\sum_{i=1}^{n} {a_{i}^{2}}(t)g_{(i)}, \\
{a_{i}}(t) &=& \exp({\beta^{i}}(t)) =
A_{i}[\sinh(rt  /T)/r]^{\sigma}
[\tanh (rt /2T)/r]^{\bar{\alpha}^{i}}, \label{53} \\
\exp({\varphi}(t)) &=& \exp({\beta^{n+1}}(t)) =
A_{n+1} [\tanh (rt /2T)/r]^{\bar{\alpha}^{n+1}}, \label{54}
\end{eqnarray}
where $t > 0, r = \sqrt{\delta} = \sqrt{\Lambda/|\Lambda|}=
\sqrt{\pm 1}$, $\sigma = (D-1)^{-1}$, $A_{i} \neq 0$ are constants,
$i=1, \ldots, n$,
and the parameters $\bar{\alpha}^{I}$ satisfy the relations
\begin{eqnarray}\label{55}
&&\frac{1}{2} \sum_{I=1}^{n+1} u_{I} \bar{\alpha}^{I}=
\sum_{i=1}^{n} d_{i} \bar{\alpha}^{i} = 0, \\
&&\sum_{I,J=1}^{n+1} G_{IJ} \bar{\alpha}^{I} \bar{\alpha}^{J} =
\sum_{i=1}^{n} d_{i} (\bar{\alpha}^{i})^{2} +
(\bar{\alpha}^{n+1})^{2} = (D-2)/(D-1). \label{56}
\end{eqnarray}
The first relation (\ref{55}) can be easily proved, using the
definition (\ref{44}) and the following identity (we remind that
due to (\ref{20}) and (\ref{25})
$ u_{I} = 2q V_{I}^{0}$)
\begin{equation}\label{57}
\sum_{I=1}^{n+1} u_{I} \bar{V}^{I}_{A} =
2q \sum_{I=1}^{n+1} V_{I}^{0} \bar{V}^{I}_{A} =
 2q \delta^{0}_{A} = 0
\end{equation}
for $A > 0$. The relation (\ref{56}) follows immediately from
(\ref{11}), (\ref{27}), (\ref{31c}), (\ref{44}) and (\ref{55}).

%%%%%%%%%%%%%%%%%%%%%%%%%%%%%%%%%%%%%%%%%%%

Now we consider the case  ${{{\cal E}}} = 0$.  In this case
for $\Lambda > 0$ there exist also an exceptional solution with
the following scale factors in  (\ref{52})
\begin{equation}\label{58}
{a_{i}}(t) = \bar{A}_{i} \exp(\pm \sigma t/T).
\end{equation}
$i=1, \ldots, n$, and ${\varphi}(t) = $const. (This solution
can be readily obtained using the formulas (\ref{32}) and (\ref{34}).)

It is interesting to note that for $\Lambda > 0$ the solution
(\ref{58}) with the sign $''+''$ is an attractor for the solutions
(\ref{53}), i.e.
\begin{equation}\label{59}
{a_{i}}(t)   \sim   \bar{A}_{i} \exp( \sigma t/T), \qquad i=1, \ldots,
n,
\end{equation}
and ${\varphi}(t) \sim  $const for
$t \rightarrow +\infty$.
The relation (\ref{59}) is the isotropization condition.
We note that the solution (\ref{52})-(\ref{56})  with
$\bar{\alpha}^{n+1} = 0$
was considered previously in \cite{20}. The special case of this
solution with $n=2$ was considered earlier in \cite{21}.

The volume scale factor  corresponding to (\ref{53}) has the form
\begin{equation}\label{60}
v = \prod_{i=1}^{n} a_{i}^{d_{i}} =
(\prod_{i=1}^{n} A_{i}^{d_{i}})  \sinh(rt  /T)/r
\end{equation}
It oscillates for negative value of cosmological constant
and exponentially increases  as $t \rightarrow +\infty$ for
positive value. For positive $A_{i}$ and ${{{\cal E}}}$ the
following identity takes place
\begin{equation}\label{61}
\prod_{i=1}^{n} A_{i}^{d_{i}} = \sqrt{{{{\cal E}}} / |\Lambda|} .
\end{equation}
For small time values we have the following asymptotical
relations
\begin{equation}\label{62}
{a_{i}}(t)   \sim   c_{i} t^{\alpha_{i}}, \qquad
\exp {\varphi}(t) \sim  c_{n+1} t^{\alpha_{n+1}}
\end{equation}
as $t \rightarrow 0$, $i=1, \ldots ,n$, where
\begin{equation}\label{63}
\alpha_{i} = \bar{\alpha}^{i}   + \sigma, \qquad
\alpha_{n+1} = \bar{\alpha}^{n+1},
\end{equation}
are Kasner-like parameters, satisfying (see (\ref{55}), (\ref{56})) the
relations
\begin{equation}\label{64}
\sum_{i=1}^{n} d_{i} \alpha_{i} =
\sum_{i=1}^{n} d_{i} (\alpha_{i})^{2} +
\alpha_{n+1}^{2} = 1.
\end{equation}
The behaviour of the scale factors near
the singularity coincides with that for the case $\Lambda = 0$
\cite{22,23}. For $\alpha_{n+1}=0$ see also \cite{11a}.

We note also that in terms of  $\alpha_{i}$-parameters
the solution (\ref{52}) - (\ref{56}) reads
\begin{eqnarray}\label{65}
{a_{i}}(t) &=&
\bar{A}_{i}[\sinh(rt /2T)/r]^{\alpha_{i}}
[\cosh (rt /2T)]^{2\sigma - \alpha_{i}}, \\
\exp({\varphi}(t)) &=&
A_{n+1} [\tanh (rt /2T)/r]^{\alpha_{n+1}} . \label{66}
\end{eqnarray}
Let us apply these solutions to the case of two spaces $(n=2)$.
{}From (\ref{64}) and (\ref{65}) we find for the non-exceptional
solutions
\ba{65a}
{a_{1}}^{\pm} &=&
{A}_{1}[\sinh(rt  /2T)/r]^{\alpha_{1}^{\pm}}
[\cosh (rt /2T)]^{\alpha_{1}^{\mp}}, \\
{a_{2}}^{\pm} &=&
{A}_{2}[\sinh(rt  /2T)/r]^{\alpha_{2}^{\mp}}
[\cosh (rt /2T)]^{\alpha_{2}^{\pm}}, \label{65b}
\ea
where
\be{65c}
\alpha_1^{\pm} = \frac{d_1 \pm \sqrt{R}}{d_1(d_1+d_2)}, \qquad
\alpha_2^{\pm} = \frac{d_2 \pm \sqrt{R}}{d_2(d_1+d_2)},
\ee
and
\be{65d}
R = d_1d_2[(d_1+d_2)(1-\alpha_3^2)-1].
\ee
Graphically these solutions are presented in Figs. 1-2.

In the Euclidean case after the Wick rotation $(t \rightarrow -it)$
we get the following instanton solutions
\begin{eqnarray}\label{67}
g &=& dt \otimes dt +\sum_{i=1}^{n} {a_{i}^{2}}(t)g_{(i)}, \\
{a_{i}}(t) &=&
\tilde{A}_{i}[\sinh(t s /2T)/s]^{\alpha_{i}}
[\cosh (t s/2T)]^{2\sigma - \alpha_{i}}, \label{68} \\
\exp({\varphi}(t)) &=&
\tilde A_{n+1} [\tanh (t s/2T)/s]^{\alpha_{n+1}} .  \label{69}
\end{eqnarray}
where $T$ is defined by (\ref{40}), $s= \sqrt{- \Lambda/
|\Lambda|}$ and the parameters $\alpha_I$ satisfy the relations
(\ref{64}). For $\Lambda < 0$ we have the special solution
$({\cal E} = 0)$
\begin{equation}\label{70}
{a_{i}}(t) = \bar{A}_{i} \exp(\pm \sigma t/T).
\end{equation}
We note that for the Euclidean case the scale factors
may be obtained from the corresponding Lorentzian
ones by the substitution $\Lambda  \mapsto - \Lambda $.
For $n=3, ~d_1=d_2=d_3=1, ~\Lambda > 0$ the special case of the solution
(\ref{67}) - (\ref{69}) was considered in \cite{24}. We note that for
$\Lambda < 0$ there are wormhole-like sections of the total metric
(\ref{67}). This takes place, for example, if $n=2, ~\alpha_3^2 \le
1-d_2^{-1},~1<d_1<d_2$, (see Fig. 2). In this case the scalar field is
real in Euclidean region.

Now, we consider the solutions of the field equations with complex
scalar field and real metric. In this case ${{{\cal E}}}, p^{1}, \ldots,
p^{n-1}$ are real and hence (see (\ref{31c})) $p^{n}$  is either real
or pure imaginary. The case of real $p^{n}$ was considered above.

For pure imaginary $p_{n}$ we have three subcases:
a) ${{{\cal E}}} >0$, b) ${{{\cal E}}} = 0$, c) ${{{\cal E}}} < 0$.
In the first case a) ${{{\cal E}}} > 0$
after the reparametrization (\ref{39}), (\ref{40}) we get the solutions
(\ref{53})-(\ref{56}) with an imaginary value of
${\bar\alpha}^{n+1}$. The cases  b) and c): ${{{\cal E}}} \leq 0$ take
place only for $\Lambda > 0$.

Let us consider the case c)  ${{{\cal E}}} < 0$.
Here, we have (see (\ref{46}), (\ref{47})) imaginary
${\bar\alpha}^{k}$:
\begin{equation}\label{456}
{\bar\alpha}^{k} = i ~\sigma_{k},
\qquad k =1, \ldots, n, ~{\bar\alpha}^{n+1} =
\sigma_{n+1}.
\end{equation}
The solution may be obtained from
(\ref{53})-(\ref{56}) substituting (\ref{456}) and $t/T \mapsto t/T
+ i\frac{\pi}{2}$:
\begin{eqnarray}\label{457}
g &=&- dt \otimes dt +\sum_{i=1}^{n} {a_{i}^{2}}(t)g_{(i)}, \\
{a_{i}}(t) &=& \hat{A}_{i}[\cosh(t/T)]^{\sigma} [{f}(t/2T)]^{\sigma_{i}},
\label{458}\\
{\varphi}(t) &=& c + 2i \sigma_{n+1} \arctan e^{-t/T}, \label{459}
\end{eqnarray}
where $c, \hat{A}_{i} \neq 0$ are
constants, $i =1, \ldots ,n$, $\sigma = (D-1)^{-1}$, $T$ is defined
in (\ref{40}), $\Lambda > 0$ and the real parameters $\sigma_{I}$ satisfy
the relations
\begin{eqnarray}\label{460}
 \sum_{i=1}^{n} d_{i}
\sigma_{i} &=& 0, \\
- \sum_{i=1}^{n} d_{i} \sigma_{i}^{2} + \sigma_{n+1}^{2} &=&
(D-2)/(D-1).  \label{461}
\end{eqnarray}
Here
\begin{equation}\label{462}
{f}(x) \equiv [\tanh(x + i \frac{\pi}{4})]^{i} = \exp(-2 \arctan e^{-2x})
\end{equation}
is smooth monotonically increasing function bounded by
its asymptotics: \hfil  \\$e^{-\pi} < {f}(x) < 1$;
${f}(x) \rightarrow 1$ as $x
\rightarrow + \infty$ and  ${f}(x) \rightarrow e^{-\pi}$ as $x
\rightarrow - \infty$ (see Fig. 3). The solution (\ref{457})-(\ref{461})
may be also obtained from formulas (\ref{32}), (\ref{33}).
The relation between the harmonic and the proper times
(\ref{41}) is modified for our case ${{{\cal E}}} < 0$
\begin{equation}\label{463}
\cosh(t /T)= 1/ \sin(q \sqrt{2|{\cal
E}|}(\tau - \tau_{0})).
\end{equation}
For the volume scale factor we have
\begin{equation}\label{464}
v = \prod_{i=1}^{n} a_{i}^{d_{i}} =
(\prod_{i=1}^{n} \hat{A}_{i}^{d_{i}})  \cosh(t  /T).
\end{equation}
The scalar field varies ${\varphi}(t)$ varies from $c +  i \pi
\sigma_{n+1}$  to $c$ as  $t$  varies from $- \infty$  to $+ \infty$. The
solution  (\ref{457})-(\ref{461})  is non-singular for $t \in (-
\infty ,+ \infty)$. Any scale factor  ${a_{i}}(t)$ has a minimum for
some $t_{0i}$ and
\begin{equation}\label{465}
{a_{i}}(t) \sim A_{i}^{\pm} \exp(\sigma |t| /T),
\end{equation}
for $t \rightarrow \pm \infty$.

The  Lorentzian solutions considered above have also
Euclidean analogues for \hfill \\ $\Lambda<0$
\begin{eqnarray}\label{466}
&&g= dt \otimes dt +\sum_{i=1}^{n} {a_{i}^{2}}(t)g_{(i)}, \\
&&{a_{i}}(t) = \hat{A}_{i}[\cosh(t /T)]^{\sigma}
[{f}(t/2T)]^{\sigma_{i}}, \label{467}  \\
&&{\varphi}(t) = c + 2i \sigma_{n+1} \arctan e^{-t/T}, \label{468}
\end{eqnarray}
with the parameters $\sigma_{I}$ satisfying  the relations
(\ref{460})-(\ref{461}). This solution may be interpreted
as classical Euclidean wormhole solution. An interesting
special case of solution (\ref{466})-(\ref{468}) occurs for
$\sigma_{i} = 0$, $i =1, \ldots ,n$, (this corresponds to $p^i = 0$)
\begin{eqnarray}\label{469}
{a_{i}}(t) &=& \hat{A}_{i}[\cosh(t /T)]^{\sigma}, \\
{\varphi}(t) &=& c \pm 2i q^{-1} \arctan e^{-t/T}. \label{470}
\end{eqnarray}
All scale factors (\ref{469}) have a minimum at $t = 0$ and
are symmetric with respect to time inversion:
$t \mapsto -t$. We want to stress here that wormhole soluions take
place only in the presence of an imaginary scalar field in the
Euclidean region. Analytic continuation of the solutions (\ref{469}),
(\ref{470}) into the Lorentzian region leads to real geometry and real
scalar field there.

%%%%%%%%%%%%%%%%%%%%%%%%%%%%%%%%%%%%%%%%%%%%%%%%%%%%%%%%%%

\subsection{Quantum wormholes}
%%%%%%%%%%%%%%%%%%%%%%%%%%%%%%%%%%%%%%%%%%%%%%%%%%%%%%%%%%%%%%%%%%%%%%%%%%%

The model introduced above leads to the WDW equation (\ref{22})
%25
\begin{equation}\label{125}
- 2 \hat H \Psi \equiv \left[ - \frac{\partial}{{\partial} z^{0}}
\frac{\partial}{{\partial} z^{0}}
+ \sum_{i=1}^{n} \frac{\partial}{\partial{}z^{i}}
\frac{\partial}{\partial{}z^{i}} - 2 \Lambda \exp(2qz^{0})\right]\Psi=0.
\end{equation}
We are seeking the solution of (\ref{125}) in the form
%26
\be{126}
{\Psi}(z)=\exp(i \vec{p} \vec{z}){\Phi}(z^{0}),
\end{equation}
where $\vec{p} = (p^{1}, \ldots , p^{n})$ is a constant vector
(generally from C$^{n}$), $\vec{z}=(z^{1}, \ldots, z^{n-1}, z^{n}=\varphi )$,
$\vec{p}\vec{z} \equiv \sum_{i=1}^n p_{i}z^{i}$ and $p_i =
\sum_{j=1}^n\eta_{ij}p^j = p^i$.
The substitution of (\ref{126}) into (\ref{125}) gives
%27
\be{127}
[- \frac12 (\frac{\partial}{\partial z^{0}})^{2} +V_0(z^{0})]
\Phi = {{\cal E}} \Phi,
\end{equation}
where
${{\cal E}}=\frac12\vec{p}\vec{p}$ and
$V_0(z^0) = - \Lambda e^{2qz^0}$. The potential $V_0(z^0)$ is plotted
on fig. 4 and fig. 5 for $\Lambda > 0$ and $\Lambda < 0$ respectively.
The clasiically allowed (Lorentzian) and forbidden (Euclidean) regions
are shown there with respect to the energy levels ${{\cal E}}$.
Solving (\ref{127}),
we get
%28
\be{128}
{\Phi}(z^{0})={B_{i\sqrt{2{{\cal E}}}/q}}(\sqrt{-2\Lambda}
q^{-1} e^{qz^{0}}),
\end{equation}
where $i \sqrt{2{{\cal E}}}/q= i|\vec{p}|/q,$ and $B=I, K$
are modified Bessel functions. We note, that
%29
\be{129}
v=\exp{qz^{0}}=\prod_{i=1}^{n}a_{i}^{d_{i}}
\end{equation}
is proportional to the spatial volume of the universe.

The general solution of Eq. (\ref{125}) has the following form
%30
\be{130}
{\Psi}(z) =\sum_{B=I,K} \int d^n \vec{p}
{~C_{B}}(\vec{p}) e^{i\vec{p}\vec{z}}{B_{i|\vec{p}|/q}}
(\sqrt{-2\Lambda} q^{-1} e^{qz^{0}}),
\end{equation}
where functions $C_{B}$ ($B=I,K$) belong to an
appropriate class. Similar solutions were found for the two-component
model  ($n=2$) and $\Lambda > 0$ in \cite{25}.

The solutions (\ref{126}) are the eigenstates of the
quantum-mechanical operators $\hat \Pi_{z^i} = -
(i/N)\partial/\partial z^i, i = 1, \dots, n$ with the eigenvalues
$(1/N)p_i$ where $N=1$ for the Lorentzian space-time region and $N=i$
for the Euclidean one.

Due to the well known time problem in quantum cosmology
the WDW equation is not really the Schr\"odinger equation. There is no
generally accepted procedure to overcome this problem but for our
particular model we can introduce some time coordinate into the
quantum equations in analogy to \cite{8}.

We split the WDW operator $\hat H$ (\ref{125}) into two parts
\be{131}
\hat H = - \hat H_0 + \hat H_1,
\ee
where
\be{132}
\hat H_0 = - \frac12 \frac{\partial^2}{{\partial z ^0}^2} - \Lambda
e^{2qz^0},
\ee
and
\be{133}
\hat H_1 = - \frac12 \sum_{i=1}^n \frac{\partial^2}{{\partial z^i}^2}.
\ee
Then the WDW equation (\ref{125}) becomes
\be{134}
\hat H_0 \Psi = \hat H_1 \Psi.
\ee
Applying $\hat H_1$ to the wave function (\ref{126}) one gets
\be{135}
\hat H_1 \Psi = {{\cal E}} \Psi.
\ee
Now, we take ${{\cal E}}$ to be real. Then,  equation (\ref{135})
shows that ${{\cal E}}$ can be treated as the energy of the subsystem
$\hat H_1$ and equation (\ref{135}) becomes the Schr\"odinger
equation. From this point of view $\Psi$ gives the stationary
states of the subsystem described by the wave equation (in Lorentzian region)
\be{136}
i \frac{\partial \tilde\Psi}{\partial \tau}  = \hat H_1 \tilde\Psi,
\ee
where
\be{137}
\tilde\Psi = e^{-i{{\cal E}} \tau} \Psi.
\ee
It can be easily seen that the wave equation
\be{138}
i \frac{\partial \tilde\Psi}{\partial \tau}  = \hat H_0 \tilde\Psi
\ee
is reduced to equation (\ref{127}).

In the semiclassical limit for the wave function (\ref{137}) equations
(\ref{136}) and (\ref{138}) are reduced to the classical equations
(\ref{31a}),  (\ref{31b}).  Indeed, the wave function (\ref{137})
can be rewritten in the form
\be{139}
\tilde\Psi = e^{-i{{\cal E}}\tau}e^{iS_1} \Phi(z^0)
\ee
with $S_1 = \sum_{i=1}^n p_iz^i$. In the semiclassical limit
the wave function $\Phi(z^0)$ takes the form
\be{140}
\Phi(z^0) = C(z^0) e^{iS_0}
\ee
with $C(z^0)$ being a slowly varying function and $S_0(z^0)$ being a
rapidly varying phase. Time is defined in the
semiclassical limit as an affine parameter along integral curves
\be{141}
\frac{\partial}{\partial\tau} = \sum_{i=0}^n \frac{\partial (S_0 +
S_1)}{\partial z_i}\frac{\partial}{\partial z^i}
\ee
where $z_0 = - z^0, ~z_i = z^i, ~i= 1, \dots, n,$
%or $z_i = \eta_{ij}z^j$.
As result we find the equations $\dot z^i = p^i, ~i= 1,
\dots, n,$ and these  coincide with the classical equations
(\ref{31a}). For this reason we used the same notation for the
constants of integration in (\ref{31a}) and for the momenta in the
wave function
%%%%%% to be corrected
(\ref{126}).
%%%%%%%%%%%%%%%%%%%%%%%%%%%%%%%
The velocity along $z^0$ is found to be $\dot z_0 = - \dot z^0 =
\partial S_0/\partial z^0$. Using this relation and
putting the wave function (\ref{139}), (\ref{140}) into equation
(\ref{138}) we reproduce the classical equation (\ref{31b}).

As shown above the parameter ${{\cal E}}$ can be interpreted as energy.
So we may treat the state ${{\cal E}} = 0$ as the ground state of the
system. The demand of reality of the geometry leads to real momenta
$p^i ~(i = 1, \dots, n-1)$  in the Lorentzian region. The scalar field
can be real or imaginary there.
In the ground state we put all momenta $p^i ~(i =
1, \dots, n)$ equal to zero and the ground state wave function reads
\be{142}
\Psi_0 = B_0 \left(q^{-1} \sqrt{-2\Lambda}  e^{qz^0} \right).
\ee
It is interesting to note that $\Psi_0$ is invariant with respect to
the rotation group O(n) in the space of vectors $\vec z = (z^1, \dots,
z^n)$.

In eq. (\ref{142}) $B_0$ denotes the Bessel functions of order zero.
A particular solution may be specified by boundary conditions. For
example, quantum wormhole boundary conditions were presented in the
Introduction. Among the  different types of boundary conditions for
wave functions describing the universe the most popular is the
Hartle-Hawking (HH) boundary condition \cite{1}. According to the HH
proposal the ground-state wave function of the universe $\Psi^{HH}_0$
is given by a path integral over all compact Euclidean geometries and
the regular matter fields:
\be{143}
\Psi_0^{(HH)} = \int d[g] d[\varphi] e^{-I_E},
\ee
where $I_E$ is the Euclidean action. For our model the Euclidean
action in harmonic time gauge and in z-coordinates reads
\be{144}
I_E = \left. \frac12 \int_{\tau^*}^\tau d \tau \left[ - (\dot z^0)^2 +
\sum_{i=1}^n (\dot z^i)^2 + 2 \Lambda e^{2qz^0}\right] - \frac12
\frac{\dot v}{v} \right|_{\tau^*}
\ee
where $v$ denotes the spacial volume
%%%%%%%%
(\ref{129})
%%%%%%%%%%%
up to a numerical factor. The upper limit $\tau$ corresponds to the
boundary of the D-dimensional  manifold, where the $z^i ~(i = 0, \dots, n)$
have values indicated by the arguments of $\Psi_0^{(HH)}$. The lower
limit $\tau^*$ corresponds to the point where the D-dimensional
manifold closes in a smooth way. The origin of the second term in
(\ref{144}) was explained in detail, e.g. by Louko \cite{17}. In the
semiclassical limit the wave function is given by
\be{145}
\Psi_0^{(HH)} \approx e^{-I_E^{cl}},
\ee
where $I_E^{cl}$ should be calculated on the classical Euclidean
solutions with boundary conditions defined by the concrete scheme of
geometry closing at $\tau = \tau^*$.

Now, we find the relationship between the HH wave function (\ref{143}),
(\ref{145}) and our ground-state wave functions (\ref{142}). Let us
first consider the case of a negative cosmological constant $\Lambda <
0$. Then, we have the  classical Euclidean equations
\be{146}
\dot z^0 = \pm \sqrt{2|\Lambda|} e^{qz^0}.
\ee
The spacial volume $v$ may be presented in the form
\be{147}
v = e^{qz^0} = - \left( q \sqrt{2|\Lambda|} ~\tau \right)^{-1}, \qquad
-\infty < \tau < 0.
\ee
Formula (\ref{147}) shows that the geometry closes at the harmonic
time $\tau \rightarrow - \infty$. It is easy to see from (\ref{147})
that the second term in (\ref{144})  contributes nothing  to $I_E$.
So, on these classical solutions the Euclidean action $I_E^{cl}$ reads
\be{148}
I_E^{cl} = \frac{1}{q^2}\frac1\tau = \frac{- \sqrt{2|\Lambda|}}{q}
e^{qz^0}
\ee
and we get the semiclassical HH wave function (\ref{145})
\be{149}
\Psi_0^{(HH)} \approx \exp\left( \frac{\sqrt{2|\Lambda|}}{q}
e^{qz^0} \right) = \exp\left( \frac{\sqrt{2|\Lambda|}}{q}
\prod_{i=1}^n a_i^{d_i} \right).
\ee
Eqn. (\ref{149}) shows that (for the class of real Euclidean
geometries)
\be{150}
\Psi_0^{(HH)} {\rightarrow} +
\infty, \qquad z^0 \rightarrow + \infty.
\ee
This conditions provides us with the possibility to chose a
solution of equation
%%%%%%%%%%
(\ref{125})
%%%%%%%%%%
corresponding to the HH ground state:
\be{151}
\Psi_0^{(HH)} = I_0 \left( \frac{\sqrt{2|\Lambda|}}{q}
e^{qz^0} \right).
\ee
The vacuum solution (\ref{151}) has the asymptotic form
\be{152}
\Psi_0^{(HH)} {\rightarrow}
\exp\left(\frac{\sqrt{2|\Lambda|}}{q} e^{qz^0} \right), \qquad
z^0 \rightarrow + \infty,
\ee
which coincides with (\ref{149}).

A similar procedure can be performed for a positive cosmological
constant $\Lambda > 0$ (see, e.g. \cite{8}). In this case the
classical Euclidean equation is
\be{153}
\left(\frac{\dot v}{v} \right)^2 + 2 q^2 \Lambda  v^2 = 0
\ee
and gives an imaginary geometry. This reflects the fact that the
geometry should be purely Lorentzian in the case $\Lambda > 0$ for
${{\cal E}} \ge 0$. The action $I_E^{cl}$ (\ref{144}) is indefinite in
this case. We can avoid this problem if we perform the analytic
continuation $v \rightarrow iv$. After this continuation the action
(\ref{144}) is formally reduced to the action in the case $\Lambda <
0$. Again, it leads to the following asymptotic for the HH wave
function:
$\Psi_0^{(HH)} {\rightarrow}
+ \infty, ~{v \rightarrow + \infty}$, which shows that the wave function
$I_0[(\sqrt{2\Lambda}/q)v]$ is a solution  of eq.
%%%%%%%%%%%%%%
(\ref{125})
%%%%%%%%%%%%%%%%%%%%%%%%%
corresponding to the HH ground state for the class of Euclidean
solutions considered here. Thus, after analytic
continuation to real values of $v$ the vacuum state corresponding to
the HH boundary condition is
\be{154}
\Psi_0^{(HH)} = J_0 \left(\frac{\sqrt{2\Lambda}}{q}v \right).
\ee
This solution has the asymptotic form
\be{155}
\Psi_0^{(HH)} {\rightarrow}
\cos\left(\frac{\sqrt{2\Lambda}}{q}v \right) =
\cos\left(\frac{\sqrt{2\Lambda}}{q}\prod_{i=1}^n a_i^{d_i} \right),
\qquad {v \rightarrow + \infty}.
\ee
For the Bianchi I universe $(n = 3, d_1 = d_2 =d_3 = 1)$ eq.
(\ref{155}) is reduced to
\be{156}
\Psi_0^{(HH)} {\sim}
\cos\left(2\sqrt{\Lambda/3}a_1a_2a_3 \right), \qquad v \rightarrow + \infty.
\ee
Similar results for the Bianchi I universe were obtained earlier in
papers by Laflamme and Louko \cite{17}.

Now, let us turn to quantum wormholes. We restrict our consideration
to real values of $p_{i}$. This corresponds to real geometries in the
Lorentzian region. In this case we have  ${{\cal E}} \geq 0$.

If $\Lambda >0$ the wave function $\Psi$ (\ref{126}) is not
exponentially damped when $v \rightarrow \infty$, i.e. the condition
(i) for quantum wormholes (see the Introduction) is not satisfied. It
oscillates and may be interpreted as corresponding to the  classical
Lorentzian solution.

For $\Lambda<0$, the wave function (\ref{126}) is exponentially damped for
large  $v$ only, when $B=K$ in (\ref{128}). But in this case the function
$\Phi$ oscillates an infinite number of times, when
$v \rightarrow 0$. So, the condition (ii) is not satisfied. The
wave function describes the transition between Lorentzian and
Euclidean regions.

The functions
%31
\be{231}
{\Psi_{\vec{p}}}(z)=
e^{i\vec{p}\vec{z}}{K_{i|\vec{p}|/q}}
(\sqrt{-2\Lambda} q^{-1} e^{qz^{0}}),
\end{equation}
may be used for constructing quantum wormhole solutions.
Like in \cite{9,10,18} we consider the superpositions of singular solutions
%32
\be{232}
{\hat{\Psi}_{\lambda,\vec{n}}}(z)=\frac{1}{\pi} \int_{-\infty}^
{+\infty} dk {\Psi_{qk \vec{n}}}(z)e^{-ik \lambda},
\end{equation}
where $\lambda \in \R$, $\vec{n}$ is a unit vector
$(\vec{n}^{2}=1)$
%($\vec{n} \in S^{n-1}$),
and the quantum
number $k$ is connected with the quantum number
${{\cal E}} = \frac12 \mid \vec p \mid^{2}$ by the formula
$2{{\cal E}} = q^{2} k^{2}$.
The calculation gives
%33
\be{233}
{\hat{\Psi}_{\lambda,\vec{n}}}(z)
=exp[-\frac
{\sqrt{-2\Lambda}}{q}e^{qz^{0}} \cosh(\lambda-q\vec{z}\vec{n})].
\end{equation}
It is not difficult to verify that the formula (\ref{233}) leads to
solutions of the WDW equation (\ref{125}), satisfying the quantum
wormholes boundary conditions.

We also note that the functions
%34
\be{234}
\Psi_{m,\vec{n}}={H_{m}}(x^{0}){H_{m}}(x^{1})\exp[-\frac
{(x^{0})^{2}+(x^{1})^{2}}{2}] ,
\end{equation}
where
%35
\ba{235}
&&x^{0}=(2/q)^{1/2}(-2 \Lambda)^{1/4} \exp(qz^{0}/2) \cosh(
\frac{1}{2}q \vec{z}\vec{n}),  \\
%236
&&x^{1}=(2/q)^{1/2}(-2 \Lambda)^{1/4} \exp(qz^{0}/2) \sinh(
\frac{1}{2}q \vec{z}\vec{n}),
\end{eqnarray}
$m=0, 1, \ldots, $ are also solutions of the WDW equation
with the quantum wormhole boundary conditions. Solutions of such
type were previously considered in \cite{5,9,10}. They are called
discrete spectrum quantum wormholes.

%These results can be easily generalized to the case of a massless scalar
%field ($\tilde V(\beta, \varphi) = 0$) minimally coupled to gravity.
%If we define $z^{n} = \varphi$, then all formulas of
%this subsection are valid with the substitution
%$n \mapsto n+1$.

It is clear from the equation (\ref{127}) and fig. 4 that in the case
$\Lambda < 0$ a Lorentzian region exists as well as an Euclidean one
for ${\cal E} > 0$. In the case $\Lambda > 0$ only the Lorentzian
region occurs for ${\cal E} \ge 0$ and for ${\cal E} < 0$ both of
these regions exist (see fig. 5). The condition ${\cal E} < 0$ leads
for pure gravity to a complex geometry in the Lorentzian region. We
can avoid this problem by the help of a free scalar field, because in
this case $2 {\cal E} = \sum_{i=1}^{n - 1} p_{i}^{2} + p_{n}^{2}$ and
we can achieve ${\cal E}< 0$ for real $p_{i} (i = 1, \dots, n - 1)$
and imaginary $p_{n}$, i.e. for an imaginary scalar field in the
Lorentzian region. The wave functions (\ref{126}), (\ref{128}) with
$\Lambda > 0$ and ${\cal E} < 0$ describe the transitions between
Euclidean and Lorentzian regions, i.e. tunneling universes.

%%%%%%%%%%%%%%%%%%%%%%%%%%%%%%%%%%%%%%%%%%%%%%%%%%%%%%%%%%%%%%%%%%%%
%%%%%%%%%%%%%%%%%%%%%%%%%%%%%%%%%%%%%%%%%%%%%%%%%%%%%%%%%%%%%%%%%%%%
\section{CLASSICAL WORMHOLES, \newline FINE TUNING OF PARAMETERS}
\setcounter{equation}{0}
%%%%%%%%%%%%%%%%%%%%%%%%%%%%%%%%%%%%%%%%%%%%%%%%%%%%%%%%%%%%%%%%%%%%

Classical wormhole solutions exist in our
model also for another interesting case. This is the case with
spontaneous compactification of the
internal dimensions. Let the factor space $M_{1}$ be our dynamical
external space. All the other factor spaces $M_{i} (i = 2, \dots, n)$
are considered as internal and static. They should be compact and the
internal dimensions have the size of order of Planck's length $L_{PL}
\sim 10^{-33}$ cm. The scale factors of the internal factor spaces
should
be constant: $a_{i} = e^{\beta^{i}} \equiv a_{(0)i} {(i = 2, \dots, n)}$.
It is not difficult to show that in the case of fine tuning of the
parameters due to
%37
\be{237}
\frac{\theta_{i}}{d_{i}a_{(0)i}^{2}} =
\frac{2 \Lambda}{D - 2} \equiv
C_{0}, \qquad i = 2, \dots, n,
\ee
all dynamical equations (\ref{14}) are reduced to one for the scale factor
$a_1=e^{\beta ^1}$ and this equation reads
%38
\be{238}
\ddot \beta ^1=-e^{2\sum\limits_1^nd_k\beta ^k}\left[ \frac{\theta _1}{d_1}
e^{-2\beta ^1}-C_0-\frac 1{d_1-1}\left( \frac 1{d_1}\frac{\partial
\tilde U}{\partial \beta ^1}+2\tilde U\right) \right] .
\end{equation}
The constraint (\ref{16}) has form
$$
d_1\left( d_1-1\right) \dot \beta ^{1^2}=\dot \varphi ^2-\theta _1e^{2\left(
d_1-1\right) \beta ^1}e^{2\sum\limits_2^nd_k\beta ^k} + e^{2d_1\beta
^1}e^{2\sum\limits_2^nd_k\beta ^k}\left[ 2\tilde U+\right.
$$
%240
\be{240}
\left. +C_0(d_1-1)\right] .
\end{equation}

{}From the equations (\ref{237}) it follows that all internal spaces
should be non-Ricci-flat and sign$\theta_{i} =$ sign$\Lambda, ~(i = 2,
\dots, n)$. We remind here that overdot denotes differentiation with
respect to the harmonic time $\tau$ \cite{7}. The minimally coupled
scalar field has the specific potential $\tilde U(\beta, \varphi) =
U(\varphi) \exp \left[-2 \sum_{i = 1}^{n}  d_{i} \beta^{i} \right]$.
For this potential it is easy to get the first integral of equation
(\ref{15})
%41
\be{241}
{\dot \varphi}^2+2U(\varphi )=\nu ^2\ =\ const .
\end{equation}
This gives
%42
\be{242}
\varphi = \nu \tau \ + \ const
\end{equation}
for $U(\varphi ) = 0$ and
%43
\be{243}
\varphi = \varphi_0\cos {m(\tau -\tau_0)}
\end{equation}
for $U(\varphi ) = \frac {m^2\varphi ^2}{2}$, where $\nu = m\varphi_0$
here.

Let us investigate the model where our
external space $M_1$ is Ricci-flat, i.e. $\theta_1=0$. Then we can
rewrite the equation (\ref{240}) as follows
%44
\be{244}
({\dot \beta}^{1})^{2}={\tilde \nu}^2+\tilde \Lambda e^{2d_1\beta^1} ,
\end{equation}
where the constants are
%45
\be{245}
{\tilde \nu}^2 = \frac {\nu^2}{d_1(d_1-1)}
\end{equation}
and
%46
\be{246}
\tilde \Lambda =\frac {2\Lambda }{d_1 \left(\sum_{k=1}^n {d_k}-1\right)}
\prod_{k=2}^n {a_{(0)k}^{2d_k}} .
\end{equation}

It is clear from equation (\ref{244}) that the dynamical behavior of
the scale factor $a_1$ depends on the signs of $\nu^{2}$ and $\Lambda$
. If $\Lambda >0$ and $\nu^2\geq 0$ then $a_1$  expands {}from zero to
infinity. For $\Lambda >0$ and $\nu^2<0$, $a_{1}$ has the turning
point at some minimum and this case may be realized for an imaginary
scalar field in the Lorentzian region. For a real scalar field in the
Lorentzian region ( i.e. $\nu^2>0$ ) and $\Lambda<0$ the scale factor
$a_1$ expands from zero to its maximum and after the turning point
shrinks again to zero. For the latter case the solution has a
continuation into the Euclidean region with the topology of a
wormhole, that means, two asymptotic regions which are connected with
each other through a throat.

Let us investigate the case with $\Lambda<0$ in more
details. As sign$\Lambda =$ sign$\theta_i\quad (i=2,\ldots,n)$
then for $\Lambda<0$ the curvatures $\theta_i<0$ also. (As a special
case the
internal spaces $M_i$ ($i=2,\ldots,n$)
may be compact spaces of constant negative curvature \cite{19}.)
The solution of equation (\ref{244}) (the Lorentzian region) has the form
%47
\be{247}
a_1(\tau )={\left[{\tilde\nu}^2/|\tilde\Lambda |\right]}^{1/2d_1}
{\left[\cosh {d_1\tilde\nu\tau}\right]}^{-1/d_1}\ ,
\qquad -\infty< \tau < +\infty .
\end{equation}

The synchronous time $t$ and the
harmonic time $\tau$ are connected by the differential equation \cite
{7}
%48
\be{248}
e^{\gamma (\tau )}d\tau = dt ,
\end{equation}
where
%49
\be{249}
\gamma (\tau )=\sum_{i=1}^n {d_i\beta^i} .
\end{equation}
It is not difficult to get the connection
%50
\be{250}
\cosh (d_1\tilde \nu \tau)= {\left[\cos \left(\sqrt {C_0d_1} ~t+
\ const\right) \right]}^{-1} .
\end{equation}
With the help of this connection we obtain the expression for the scale
factor $a_1$ with respect to the synchronous time
%51
\be{251}
a_1(t)={\left[{\tilde\nu}^2/|\tilde\Lambda |\right]}^{1/2d_1}
{\left[\sin \sqrt {C_0d_1} ~t\right]}^{1/d_1}\ , \quad
0\leq t\leq \frac {\pi}{\sqrt {C_0d_1}} ,
\end{equation}
where the constant in (\ref{250}) was fixed by condition $a(t=0)=0$. For
$t\rightarrow 0$ we have $a_1\sim t^{1/d_1}$ .
Thus the external space $M_1$ has the behavior of a FRW-universe
filled with radiation for $d_1=2$ and with ultrastiff matter for $d_1=3$.
The Lorentzian metric
(\ref{1}) in synchronous time gauge reads as
%52
\be{252}
g =-dt \otimes dt + a_1^2(t)g_{(1)}+\sum_{i=2}^n {a_{(0)i}^{2} g_{(i)}}
\end{equation}
with $a_{1}$ given by (\ref{251}).

Our next step is to get the wormhole-type solution for $M_1$ in the
Euclidean region. The transition into the Euclidean space is performed
by the Wick rotation $t \rightarrow -it$. The exact form of the
transformation from the Lorentzian time $t_{L}$ to the
Euclidean ''time'' $t_{E}$ can be obtained demanding the existence of
wormholes being symmetric with respect to the throat (see Zhuk in
\cite{4}). In what follows, for the expression (\ref{251}) we should
perform the analytic continuation $t_{L} =
\frac{\pi}{2\sqrt{C_{0}d_{1}}} - it_{E}$.
This gives us
%53
\be{253}
a_1(t)={\left[{\tilde\nu}^2/|\tilde\Lambda |\right]}^{1/2d_1}
{\left[\cosh \sqrt {C_0d_1} t\right]}^{1/d_1}\ , \quad
-\infty<t< +\infty .
\end{equation}
Of course, this formula can be obtained also as a solution to the
Euclidean analog of the equations (\ref{240}), (\ref{244}) for an imaginary
scalar field in the Euclidean region.
The metric of the Euclidean region is given by
%54
\be{254}
g = dt\otimes dt + a_{1}^{2}(t)  g_{(1)} + \sum_{i = 2}^{n} a_{(0)i}^{2}
g_{(i)} ,
\ee
where $a_{1}(t)$ is described by (\ref{253}). Thus, in Euclidean space we
have two asymptotic regions $t\rightarrow \pm\infty$
connected through a throat of the size
${\left[{\tilde\nu}^2/|\tilde\Lambda |\right]}^{1/2d_1}$
and this object is a wormhole by definition.

It is clear that the Lorentzian solution (\ref{251}) and its Euclidean
analog (\ref{253}) take place only in the presence of a real scalar
field in the Lorentzian region (i.e $\nu^2>0$) or equivalently an
imaginary scalar field in the Euclidean region.

%%%%%%%%%%%%%%%%%%%%%%%%%%%%%%%%%%%%%%%%%%%%%%%%%%%%%%%%%%%%%%%%%%%%
%%%%%%%%%%%%%%%%%%%%%%%%%%%%%%%%%%%%%%%%%%%%%%%%%%%%%%%%%%%%%%%%%%%%
\section{CONCLUSIONS}
\setcounter{equation}{0}
%%%%%%%%%%%%%%%%%%%%%%%%%%%%%%%%%%%%%%%%%%%%%%%%%%%%%%%%%%%%%%%%%%%%

In this paper we investigated multidimensional cosmological models
with $n (n > 1)$ Einstein spaces $M_{i}$ in the presence
of the cosmological constant $\Lambda$ and a homogeneous minimally
coupled scalar field $\varphi(t)$ as a matter source. The problem was
to find classical and quantum  wormhole solutions. Classical wormholes
are solutions of the classical Einstein  equations
describing Riemannian metrics with two large regions joined by a throat.
Quantum wormholes are solutions of the Wheeler-DeWitt (WDW) equation
with the proper boundary conditions proposed by Hawking and Page
\cite{5}.

The model was investigated where one of the factor spaces, say
$M_{1}$, is Ricci-flat. In the case when all other factor spaces $M_{i}, i
= 2, \dots, n,$ are Ricci-flat too,  the classical Einstein as well as
the WDW equations are integrable. For a negative cosmological constant
$\Lambda < 0$ quantum wormhole solutions were constructed. These
solutions exist for pure gravity as well as for the model with a free
minimally coupled scalar field. Classical wormhole solutions exist in
the Euclidean region for $\Lambda < 0$ in the presence of an imaginary
as well as real scalar field.

Classical wormhole solutions were also obtained in models with
spontaneous compactification. In this case the Ricci-flat factor space
$M_{1}$ was considered as our external dynamical space. All other
factor spaces $M_{i}, i = 2, \dots, n$ are static with constant scale
factors $a_{(0)i} = const$ and all of them are fine tuned to each
other and to the cosmological constant: $\frac{\theta_{i}}
{d_{i}a_{(0)i}^{2}} = \frac{\theta_{k}}{d_{k}a_{(0)k}^{2}}
= \frac{2 \Lambda}{D - 2}, \qquad i, k = 2, \dots, n$.

As in the previous model, wormhole solutions exist for a negative
cosmological constant $\Lambda < 0$. But there are important
differences. Firstly, all inner spaces $M_{i}, i = 2, \dots, n,$ are
non-Ricci-flat and have negative curvature. Secondly, the wormhole
solution for the later case exists only in the presence of an
imaginary sclar field in the Euclidean region. Thirdly, it seems
hardly to be possible in the case $\theta_{2}, \dots, \theta_{n} \neq
0$ to integrate the Einstein equations as well as the WDW equation
without the demand of spontaneous compactification with fine tuning.

In models with one scale factor having a turning point (at the
minimum)
the production of the Lorentzian space-time is treated as a quantum
tunneling process \cite{31} (''birth from nothing''). The universe
appears spontaneously going through the potential barrier with size
equal to the size of the Lorentzian universe at the turning point.
In our case of multidimensional models this kind of interpretation
becomes more complicated. It follows from (\ref{458}) that the factor
spaces $M_i$ in general reach their minimum expansion positions at
different times. The ''birth from nothing'' for each factor space
takes place at a different value of time. If the difference between
these events goes to infinity the extra dimensions are in the classically
forbidden region forever. This interpretation is in the spirit of
the Rubakov-Shaposhnikov idea \cite{32} stating that extra dimensions
are unobservable because they are hidden from us by a potential
barier.
%%%%%%%%%%%%%%%%%%%%%%%%%%%%%%%%%%%%%%%%%%%%%%%%%%%%%%%%%%%%%%%%%%%%%
\bigskip

\bigskip

%\pagebreak
\noindent
{\bf ACKNOWLEDGMENT}\\
The work was sponsored by KAI e.V. Berlin through the WIP project
016659/p and partly by the Russian Ministry of Science.  A. Z.
was supported by DFG grant 436 RUS 113-7-1, V.I. and  V. M.
by DFG grant 436 RUS 113-7-2.
V. I., V. M. and A. Z. also thank the colleagues of the WIP gravitation
project group at Potsdam University for their hospitality.

%%%%%%%%%%%%%%%%%%%%%%%%%%%%%%%%%%%%%%%%%%%%%%%%%%%%%%%%%%%%%%%%%%%%%
%%%%%%%%%%%%%%%%%%%%%%%%%%%%%%%%%%%%%%%%%%%%%%%%%%%%%%%%%%%%%%%%%%%%%
\bigskip

\bigskip

\bigskip

%\newpage

\newpage
\noindent
{\bf Figures:} \\

\noindent
{\bf Fig.~1} ~The behaviour of the scale factors $a_1^{\pm}(t),
a_2^{\pm}(t)$ (see (\ref{65a}), (\ref{65b})) for
$\Lambda<0, ~n=2, 1<d_1<d_2$ and different values of the integration
constant $\alpha_3$. The left figures correspond to the sign ''+'' in
(\ref{65a}), (\ref{65b}).

\bigskip

\noindent
Fig. 1.1 $\qquad 0 \le \alpha_3^2 < 1 - \frac{1}{d_1}$

\vspace{6truecm}
\noindent
Fig. 1.2 $\qquad \alpha_3^2 = 1 - \frac{1}{d_1}$

\vspace{6truecm}
\noindent
Fig. 1.3 $\qquad 1 - \frac{1}{d_1} < \alpha_3^2 < 1 - \frac{1}{d_2}$

\vspace{6truecm}
\noindent
Fig. 1.4 $\qquad \alpha_3^2 = 1 - \frac{1}{d_2}$

\vspace{6truecm}
\noindent
Fig. 1.5 $\qquad 1 - \frac{1}{d_2} < \alpha_3^2 < 1 - \frac{1}{d_1+d_2}$

\vspace{6truecm}
\noindent
Fig. 1.6 $\qquad \alpha_3^2 = 1 - \frac{1}{d_1+d_2}$

\newpage
\noindent
{\bf Fig.~2} ~The behaviour of the scale factors $a_1^{\pm}(t),
a_2^{\pm}(t)$ (see (\ref{65a}), (\ref{65b})) for
$\Lambda>0, ~n=2, 1<d_1<d_2$ and different values of the integration
constant $\alpha_3$. The left figures correspond to the sign ''+'' in
(\ref{65a}), (\ref{65b}).

\bigskip

\noindent
Fig. 2.1 $\qquad 0 \le \alpha_3^2 < 1 - \frac{1}{d_1}$

\vspace{6truecm}
\noindent
Fig. 2.2 $\qquad \alpha_3^2 = 1 - \frac{1}{d_1}$

\vspace{6truecm}
\noindent
Fig. 2.3 $\qquad 1 - \frac{1}{d_1} < \alpha_3^2 < 1 - \frac{1}{d_2}$

\newpage
\noindent
Fig. 2.4 $\qquad \alpha_3^2 = 1 - \frac{1}{d_2}$

\vspace{6truecm}
\noindent
Fig. 2.5 $\qquad 1 - \frac{1}{d_2} < \alpha_3^2 < 1 - \frac{1}{d_1+d_2}$

\vspace{6truecm}
\noindent
Fig. 2.6 $\qquad \alpha_3^2 = 1 - \frac{1}{d_1+d_2}$

\newpage

\vspace*{8truecm}
\noindent
{\bf Fig.~3.} ~Representation of the function $f(x) = \exp(-2\arctan
e^{-2x})$.

\vspace*{8truecm}
\noindent
{\bf Fig.~4.} ~The potential $V_0(z^0)$ for $\Lambda > 0$ (solid line)
and the energy levels ${{\cal E}}$ (dashed lines). For ${{\cal E}} \ge 0$
the Lorentzian region exists, only. For ${{\cal E}} < 0$ both regions,
the Lorentzian as well as the Euclidean one exist. In this case
quantum transitions with topology changes take place (tunneling
universe).

\newpage
\vspace*{8truecm}
\noindent
{\bf Fig.~5.} ~The potential $V_0(z^0)$ for $\Lambda < 0$ (solid line)
and the energy levels ${{\cal E}}$ (dashed lines). For ${{\cal E}} \le 0$
the Euclidean region exists, only.  For ${{\cal E}} > 0$ both regions,
the Lorentzian as well as the Euclidean one exist. In this case
quantum transitions with topology changes take place (quantum
wormholes).


\begin{thebibliography}{99}
\bibitem{1} J.B. Hartle, S.W. Hawking, Phys. Rev., {\bf D28} (1983)
            2960.
\bibitem{2} S. Giddings, A. Strominger, Nucl. Phys.,
            {\bf B306}  (1988) 890.
\bibitem{3} R.S.Myers, Phys.Rev., {\bf D38}  (1988) 1327.
\bibitem{4} S.W. Hawking, Phys. Rev., {\bf D37}   (1988) 904; Mod.
            Phys. Lett., {\bf A5}  (1990) 145; Mod.
            Phys. Lett., {\bf A5}  (1990) 453. \\
            K. Lee, Phys. Rev. Lett., {\bf 61}  (1988) 263. \\
            A. Lyons, Nucl. Phys., {\bf B324}  (1989) 253. \\
            J.J. Halliwell, R. Laflamme, Class. Quant. Grav., {\bf 6}
             (1989) 1839. \\
            B.J. Keay, R. Laflamme, Phys. Rev., {\bf D40} (1989) 2118. \\
            J.D. Brown, C.P. Burgess, A. Kshirsagar, B.F. Whiting,
            J.W. York, Nucl. Phys., {\bf B328}  (1989) 213.  \\
            A. Hosoya, W. Ogura, Phys. Lett., {\bf  B325}  (1989) 117.
              \\
            P.F. Gonzalez-Diaz, Phys. Lett., {\bf B233}  (1989) 85;
            Phys. Rev. {\bf D40}  (1989) 4184; Nuovo Cimento, {\bf
            B106}  (1991) 335; Int. Journ. Mod. Phys., {\bf A7}
            (1992) 2355.   \\
            D. Coule, K. Maeda, Class. Quant. Grav., {\bf 7}
            (1990) 955. \\
            A.K. Gupta, J. Hughes, J. Preskill, M.B. Wise, Nucl.
            Phys., {\bf B333} (1990) 195.  \\
            S. Wada, Mod. Phys. Lett., {\bf A7}  (1992) 371. \\
            A. Zhuk, Phys. Lett., {\bf A176}  (1993) 176.
\bibitem{5} S. W. Hawking, D. N. Page, Phys. Rev., {\bf D42}  (1990) 2655.
\bibitem{6} D.N. Page, J. Math. Phys., {\bf 32}  (1991) 3427.  \\
            P. Gonzalez-Diaz, Nucl. Phys., {\bf B351}  (1991) 767.  \\
            S. Wada, Mod. Phys. Lett., {\bf A7}  (1992) 371.  \\
            S. P. Kim, D.N Page, Phys. Rev., {\bf D45}  (1992) R3296.
                \\
            L.J. Alty, P.D. D'Eath, Phys. Rev., {\bf D46}
            (1992) 4402.
\bibitem{7} V.D. Ivashchuk, V.N. Melnikov and A. I. Zhuk,
            Nuovo Cimento, {\bf B104}  (1989) 575.
\bibitem{8} A. Zhuk, Class. Quant. Grav., {\bf 9}  (1992) 2029.
\bibitem{9} A. Zhuk, Phys. Rev., {\bf D45}  (1992) 1192.
\bibitem{10} A.Zhuk, Sov. Journ. Nucl. Phys., {\bf 55}  (1992) 149.
\bibitem{11} A.Zhuk, Sov. Journ. Nucl. Phys., {\bf 56}  (1993) 223.
\bibitem{11a} V.D. Ivashchuk, Phys. Lett., {\bf A170}  (1992) 16.
\bibitem{11b}Y.Shen, Z.Tan, Nuovo Cim., {\bf B107}  (1992) 653.
\bibitem{12} V.D. Ivashchuk, V.N. Melnikov, Intern. J. Mod. Phys.
             1994, in press.
\bibitem{12a} S. Chakraborty, Mod. Phys. Lett., {\bf A7}  (1992) 2463.
\bibitem{13} U. Bleyer, A. Zhuk, {\em Multidimensional integrable
                cosmological models with dynamical and spontaneous
                compactification}, Preprint Free University Berlin,
                FUB-HEP/93-19; {\em Classical and quantum behaviour of
                multidimensional integrable cosmological models},
                Preprint Free University Berlin, FUB-HEP/94-1.
\bibitem{14} M.I. Kalinin, V.N. Melnikov, in: Problems of Gravitation
             and Elementary Particle Theory. Moscow, Proc. VNIIFTRI,
             {\bf 16(46)}  (1972) 43 (in Russian).
\bibitem{15} V.N. Melnikov, V.A. Reshetov, in: Abstr. VIII
             Nation. Conf. on Element. Particle Phys. (ITP, Kiev, 1971) p.117
             (in Russian).
\bibitem{16} K.P. Stanyukovich, V.N. Melnikov. {\em
             Hydrodynamics, Fields and Constants in the Theory of Gravitation}
             (Energoatomizdat, Moscow, 1983), p.256 (in Russian).
\bibitem{17b} V.N. Melnikov, G.D. Pevtsov, in: Abstr. GR-10. Padua,
             1983, p. 571; Izvestiya Vusov, fisica, 1985, N4, p. 45.
\bibitem{17} R. Laflamme, Phys. Lett. {\bf B198}  (1987) 156; PhD
             thesis, Cambridge Univ. (1988);  \\
             J. Louko, Phys. Rev., {\bf D35}  (1987) 3760; Annals of
             Physics, {\bf 181}  (1988) 318; Class. Quant. Grav. {\bf
             5} (1988) L181; \\
             R. Laflamme, E.P.S. Shellard, Phys. Rev., {\bf D35}
             (1987) 2315;   \\
             J. Halliwell, J. Louko, Phys. Rev., {\bf D42}
             (1990) 3997;  \\
             J. Uglum, Phys. Rev., {\bf D46}  (1992) 4365.  \\
             S. Chakraborty, Mod. Phys. Lett., {\bf A6}  (1991) 3123;
             Mod. Phys. Lett., {\bf A8}  (1992) 653.
\bibitem{17a} G.W. Gibbons, S.W. Hawking, Phys. Rev., {\bf D15}
             (1977) 2752.
\bibitem{17c} V.D. Ivashchuk, V.N. Melnikov, Phys. Lett., {\bf A135}
              (1989) 465.
\bibitem{18} L. Campbell, L. Garay, Phys. Lett. {\bf B254} (1991) 49.
\bibitem{19} H.V.Fagunders, Phys. Rev. Lett., {\bf 70}  (1993) 1579.
\bibitem{20} V.D. Ivashchuk, V.N. Melnikov, Theor. and Math.
             Phys.,  (1994) in press.
\bibitem{21} D. Lorenz-Petzold, Phys. Rev., {\bf D31}  (1985) 929;
             Phys. Lett., {\bf B151} (1985) 105.
\bibitem{22} U. Bleyer, M. Rainer, A. Zhuk, {\em Classical and Quantum
             Solutions of Conformally Related Multidimensional Cosmological
             Models}, Preprint Freie Universit\"at Berlin, FUB-HEP/94-3 (1994).
\bibitem{23} U. Bleyer, A. Zhuk, {\em Kasner-like, Inflationary and
             Steady State Solutions in Multidimensional Cosmology},
             in preparation.
\bibitem{24} J. Louko, P.J. Ruback, Class. Quant. Grav., {\bf 8}
             (1991) 91.
\bibitem{25} E.I. Guendelman, A.B. Kaganovich, Phys. Lett., {\bf
             B301}  (1993) 15.
\bibitem{31} A. Vilenkin, Phys. Rev., {\bf D27} (1983) 2848.
\bibitem{32} V.A. Rubakov, M.E. Shaposhnikov, Phys. Lett., {\bf B125}
             (1983) 136.
\end{thebibliography}
\end{document}